\documentclass{article}
\usepackage[utf8]{inputenc}
\usepackage{amsmath}
\usepackage[a4paper, total={6in, 8in}]{geometry}
\usepackage{blindtext}
\usepackage{setspace}
\usepackage{graphicx}

\usepackage[affil-it]{authblk}

\usepackage[natbib=true,style=numeric,sorting=none]{biblatex}
\begin{document}

\title{False negative probability in iGSR detection: \\a Bayesian approach}
\author[1,2,3]{Martín A. Onetto}
\author[4]{Edgardo Carignano}
\author[1,3]{Rodolfo G. Pregliasco}

\affil[1]{Sección Física Forense, Centro Atómico Bariloche/Comisión Nacional de Energía Atómica (CNEA), Av. E. Bustillo 9500, R8402AGP San Carlos de Bariloche, Río Negro, Argentina}
\affil[2]{Instituto Balseiro, Universidad Nacional de Cuyo/Comisión Nacional de Energía Atómica (CNEA), Av. E. Bustillo 9500, R8402AGP San Carlos de Bariloche, Río Negro, Argentina}
\affil[3]{Consejo Nacional de Investigaciones Científicas y Técnicas (CONICET), C1033AAJ Buenos Aires, Argentina, Sección Física Forense, Centro Atómico Bariloche/Comisión Nacional de Energía Atómica (CNEA), Av. E. Bustillo 9500, R8402AGP San Carlos de Bariloche, Río Negro, Argentina}
\affil[4]{Laboratorio Forense Rosario, Organismo de Investigaciones, Ministerio Público de Acusación (MPA), Lamadrid 460, S2001EBJ Rosario, Santa Fe, Argentina}
\maketitle

\begin{abstract}
{This article discusses the detection of inorganic gunshot residue (iGSR) particles through scanning electron microscopy with energy dispersive X-ray spectrometer (SEM/EDS) in the discovery step. 
We calculated the probability that all characteristic inorganic gunshot residue particles (iGSR) go undetected and their dependence on the image pixel resolution setup. 
We built and validated a iGSR particle detection model that relates particle size with equipment registers, and we applied it to 1,174 samples analyzed by a forensic science laboratory. 
Our results indicate that the probability of missing all characteristic iGSR particles is below 5\% for pixel sizes below~$0.32 \mu m^{2}$. These values indicate that pixel sizes as great as the double in area that is commonly used in laboratory casework, $0.16 \mu m^{2
}$, are effective for an initial scanning of a sample as it yields good rates of detection of characteristic particles, which might exponentially reduce laboratory workload. }
\end{abstract}

\clearpage
\section{Introduction}

Scanning electron microscopy with energy dispersive X-ray spectrometer (SEM/EDS) is the standardized procedure to detect particles of inorganic gunshot residue (iGSR) in samples collected from hands, faces, clothes or other surfaces from an individual suspected to be near an event of a firearm discharging~\cite{WoltenI,zeichner_more_1997,ASTM2020}. iGSR particle categories as defined by ASTM E1588-20 \cite{ASTM2020} as characteristic: those with composition PbSbBa; and consistent: those with composition PbBaCaSi, BaCaSi, SbBa, PbSb,
BaAl, PbBa, Pb, Ba and Sb; commonly associated: those that have one of the following elements: Pb, Sb, Ba. Additional elements found in these particle classes are listed in \cite{ASTM2020}.

After calibration, validation and quality control are satisfied~\cite{Ritchie2020}, automated search software is the standard tool to analyze iGSR samples since year 1987~\cite{Timan1987,Kee1987}. 
As key advantages, these tool help bypass the need of a microscopy operator throughout sample scanning and the bias associated with manual particle search, hence allowing standardized and more exhaustive analysis. 
The scanning procedure comprises two distinct phases or steps: the discovery and the confirmation step~\cite{ASTM2020}. During discovery, candidate iGSR particles are identified through all the sample area using SEM equipped with a back-scatter electron detector (BSE), while in the confirmation stage these candidate particles are classified through EDS and new images are acquired.

In recent years there have been great efforts to optimize SEM/EDS configuration setup and parameters in the search for iGSR. Extensive protocols on the laboratory design, procedures, and quality control checks to avoid contamination have been discussed thoroughly in \cite{White2014}. A complete list of practices for validating the performance of the technique is depicted in \cite{Ritchie2020}, where a set of tests are provided to establish the baseline performance on the instruments used for iGSR analysis such as correct calibration of the beam current, the stage alignment, the BSE detector among others.

This work is focused on the discovery phase of iGSR particle detection. It has been found that the main parameters that affect the detection of iGSR particles are the search magnification, the BSE image acquisition time and intensity threshold. 
Although inconsistencies on the number of particles detected under optimal conditions (such as brightness and contrast among others) has been reported~\cite{Izraeli2014}, it has been mainly attributed to the BSE acquisition time per pixel and there has been no quantitative discussion on the influence and interplay of the particle sizes, their position on the sample, and the pixel size. Here we discuss the influence of the search magnification on those particles that are not detected by BSE, hereinafter referred to as “false negatives". Due to limitations on the instrumentation false negatives are inevitable however, their occurrence on the results of sample scanning can be modeled to quantify their contribution to iGSR characteristic particle final count which is the main contribution of this article.

As false negatives are iGSR particles candidates that do not enter the second step of analysis, namely the confirmation step, there is a need to minimize how often they can occur since those particles are lost in the further analysis and never recovered. The quantification of this phenomenon is relevant to understand how to keep the number of lost particles low while keeping the total run time reasonable for everyday casework. The other kind of misclassification error that can occur are false positives, which are particles that go through the first detection step but are not iGSR related. These are mostly discarded when the EDS spectra is acquired. This issue has been addressed in~\cite{Charles2020}, and the correct classification of these particles is usually well assigned in the confirmation step, without loss of iGSR information. False positive in iGSR candidates have very different consequences that false negatives. In this article we study the BSE in the first step to understand how many false negatives can be expected in real cases.

Instruments are often thought to make direct measurements of the objects of interest; however,
the sizes detected by automated microscopes are actually altered by detection strategies and there is a subtle distinction between size measurements and sizes that particles actually have. Bayesian statistic is the main framework to describe
the way in which these two things have an effect on each other.
To do this we need to build and validate a model to describe the measurement process with the BSE detector and use Bayes rule to quantitatively infer particle sizes. The interplay between particle and pixel sizes is a fundamental statistical aspect that the authors believe has not been discussed quantitatively in the literature. Bayesian methods are the appropriate standpoint to address this problem and is a growing technique in the forensic field, specifically in GSR analysis as pointed out by \cite{Charles2010}. 
We refer the reader to the pioneering work of \cite{aitken,taroni} for an introduction to the use of Bayesian statistics in forensic sciences.

Although the results of this approach won't be able to prescribe how to eliminate false negatives, it does describe how often these particles will go undetected for different image pixel sizes, as well as how probable it is to miss all characteristic iGSR particles in a sample during the discovery step which is our main concern. We will call such samples  hereinafter `false negative samples' (FNS). Throughout the article we present the results of controlled measurements to confirm the descriptive capacity of our simple model.
In Section 2 we provide 
a geometric model of particles and their microscopic detection through BSE, 
a statistical model used to infer particle size distribution 
and the calculation that yields the probability to miss all characteristic iGSR particles on a sample. 
In Section 3.1 and 3.2 we show a 
characterization and validation of the geometric model using successive measurements of the same sample with different pixel size so as to compare their particle size distribution. 
In Section 3.3 we depict the global size parametric distribution of characteristic particles 
which is fitted using a database obtained from a microscopy service unit. With these parameters, we calculate the probability that all iGSR characteristic particles on the sample go undetected, namely reporting a false negatives sample, for different pixel size settings $p_{x}$ at the end of Section~3. We found that pixel sizes as great as double the usual laboratory casework can yield good rates of detection of characteristic particles in gunshot residue samples, which may help reduce laboratory workload.

\section{Materials and methods}
\subsection{Circular particle model}
\label{secCIRC}

The automated micro-analysis system\footnote{for instance {\em INCA Energy}, the widespread standard in Argentina.} requires the operator to select search parameters, the most critical ones being the threshold value –determined by measuring a reference sample– and the smallest expected particle width. 
A pixel size is then defined which is usually half the smallest expected
particle width, and the system only detects pixels whose signal is above the threshold. Activated
pixels correspond to chemical elements with high atomic numbers, which are of special interest
in the search for particles containing heavy metals such as those in iGSR. With this measurement strategy, as we are only interested in characteristic iGSR particles we will consider as a simplification in our model of detection that a particle which area is not fully registered because partially covered pixels are disregarded.\footnote{We are aware that this simplification does not take into account the brightness differences that may exist among other relevant iGSR particles. However, since our focus is solely on detecting characteristic particles, we believe and demonstrate that this assumption in the detection model is valid.} So, instead of $A\,$, an eroded version of the particle is registered with a seemingly $B\,$ area.

Activated pixels correspond to chemical elements with high atomic numbers, which are of particular interest in the search for particles containing heavy metals, such as those found in characteristic iGSR particles. The measurement strategy focuses on detecting characteristic iGSR particles with a specific composition of PbSbBa. As a simplification in our detection model, we assume that partially covered pixels do not register the full area of a particle.
 
\begin{figure}[tbh]
\centering
\includegraphics[width=0.6\textwidth]{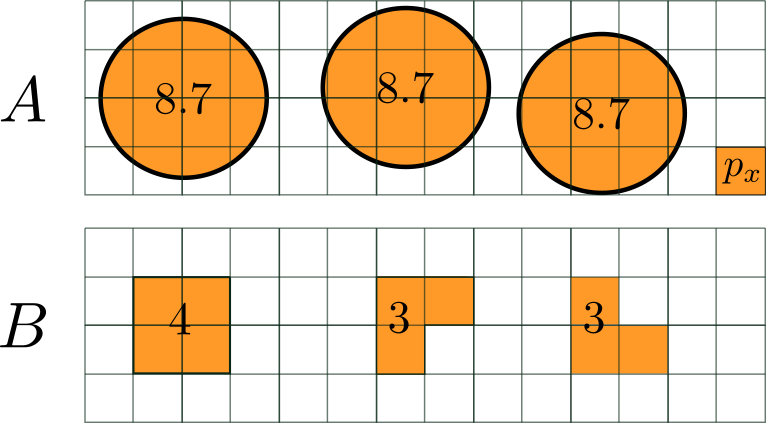}
\caption{\label{fig:dibu} Circular particle size registration. The numbers indicate the area of each particle in unit pixels. The real areas $A$ are different from the registered areas $B$. The values of $B$ are the areas occupied by completely covered pixels and depend on particle location on the grid.}
\end{figure}
In this context, we modeled the relationship between real areas $A$ and registers $B\,$, on the
hypothesis of circular particle shape. This representation poses the advantages of being simple,
symmetric, descriptive of iGSR, and suited to modeling requirements. Particle shape may be
disregarded when size exceeds pixel area $p_{x}$, as $B$ will then be a good representation of total
area. Instead, when $A$ is similar to the value of $ p_{x}$, both particle shape and position on the pixel grid affect registration $B$ (Fig.~\ref{fig:dibu}).

Different values of $B$ can be measured, given a particle of area A. The set of $B$ results is discrete,
finite, and depends on $p_x\,$ size. Although a single unequivocal value of $B$ cannot be established
on the basis of a particle of area $A$, the fact that all positions on the grid are equivalent allows
the calculation of $P(B|A)\,$\footnote{Notation for the probability of measuring $B$ given a certain value of $A$ \cite{Gelman}.}. However, as $B$ is actually measured in practice, the question arises as
to what values of $A$ are feasible. These values can be calculated using Bayes’ theorem.
\begin{equation}
\begin{aligned}
    P(A|B) &= \frac{P(B|A)P(A)}{P(B)}\\
     &\propto L(A|B)P(A)\,,\\
\end{aligned}
\label{eq:bayes}
\end{equation}

\noindent
where $L(A|B)$ is known as the likelihood function and equals the expression $P(B|A)$, although
analyzed as a function of $A$. Worth pointing out, $L(A|B)$ is not a probability distribution as a
function of $A$ but a function of $B$.  On the other hand, $P(A)$ is indeed a probability distribution known as
prior probability and represent the degree of certainty about the values that $A$ may take
regardless of any other values. When there are no reasons to believe that some values will be
more likely than others, this distribution is defined as an improper uniform distribution, i.e., $P(A) \propto 1$.

To calculate $L(A|B)$ we simulated circular particles with uniform area distribution and
placed them homogeneously on a grid. With the results of the simulation, we built the likelihood
function $L(A|B)$ as follows: for each discrete value $B$, we recorded its frequency for all $A$ simulated on the grid. This defines $L(A|B)$ for each possible discrete value $B$.


\subsection{Application to forensic laboratory samples}


The quantization introduced by the pixel grid raises the possibility that some iGSR particles in the sample
may go undetected, namely false negatives, which is critical to the analysis of this forensic technique. To quantify such probability in real cases, we applied our study to
iGSR samples collected and analyzed in a forensic science laboratory.

\begin{figure}[tbh]
\centering
\includegraphics[width=1\textwidth]{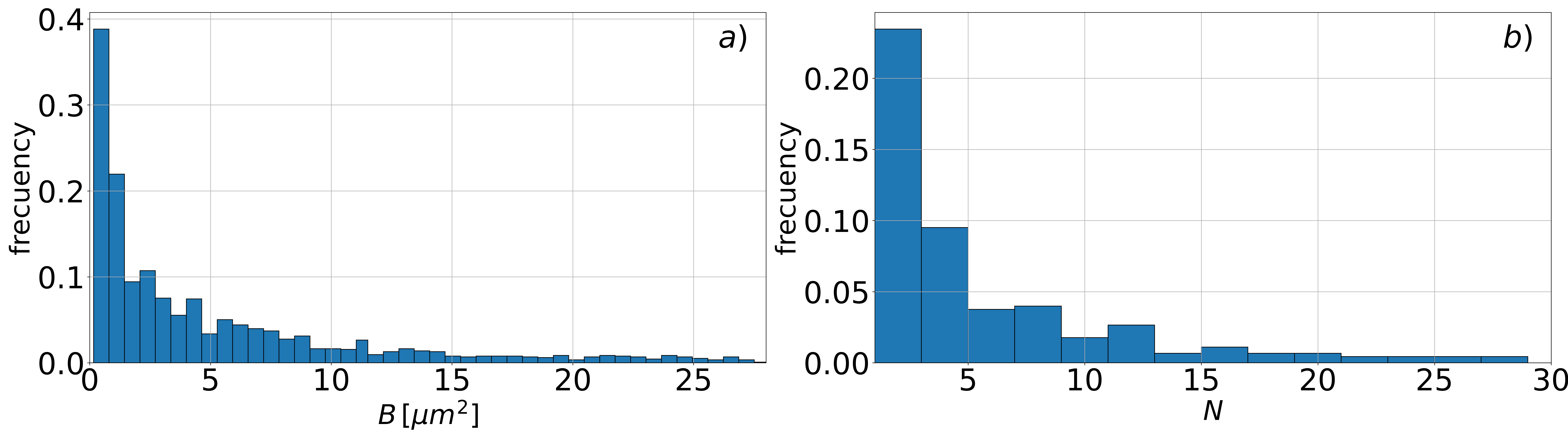}
\caption{Histograms of data recorded for the analysis of iGSR samples collected in a forensic microscopy
service unit. a) Histogram of areas $B$ recorded for characteristic particles with $p_{x}= 0.16$. b) Histogram of the number of characteristic particles, $N$.\label{fig:dist_muestras}}
\end{figure}

The use of real case samples requires a different strategy from that used in the analysis of experimental data obtained in controlled conditions. Only a few parameters (firearm, ammunition, collection time, etc.) can vary in experimental designs. For comparisons to be made with real case scenarios, however, parameters need to be weighted in a way that resembles the frequency they exhibit in practice. This entails extensive experimental work in several different conditions and the analysis of each condition’s relevance. In contrast, using all the real case samples collected at a forensic laboratory harmonizes data so that inferences can be made which are closely linked to everyday practice.
We used a set of 1,174 samples collected from hands and analyzed by Policía de Investigación de Rosario (Criminalistics Unit at Rosario Police Department, Santa Fé, Argentina) between 2017
and 2019; at least one characteristic particle was registered in 320 out of the 1,174 samples with a sum of 2069 particles.
Analyses were carried out on a Zeiss Evo LS 15 microscope (SEM) and an X-ray spectrometer (EDS) using INCA software for iGSR detection. All samples were analyzed with a resolution $p_{x} = 0.16 \mu m^{2}$ to follow ASTM requirements to detect all $1 \mu m$ diameter particles \cite{ASTM2020}. Of particular interest were the distribution of areas $B$ of characteristic iGSR particles
and the number of these particles ($N$) registered in each sample (see ~Fig.~\ref{fig:dist_muestras}).

\subsection{Distribution of A}
\label{sec:model}

\begin{figure}[tbh]
\centering
\includegraphics[width=0.8\textwidth]{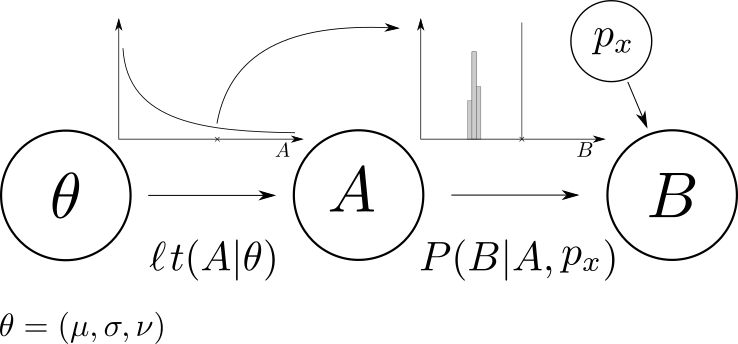}
\caption{Model diagram for the distribution of areas $A$ and their registers $B$. $A$ is described as having a $log\textrm{--}t$ Student' distribution dependent on parameters $(\mu,\,\sigma,\,\nu)$\label{fig:model_diag}. The distribution $P(B|A,p_{x})$ corresponds to the probability of obtaining a recorded value $B$, given a real size $A$ and a pixel size $p_{x}$.}
\end{figure}

Calculating the probability of failing to detect all iGSR characteristic particles in a sample implies determining how many of those may go undetected due to their size. To this end, we built a model that describes real particle area A distribution on the basis of measured values $B$~(Fig. \ref{fig:model_diag}).


The distribution of $A$ must take positive values and be long tailed, as particles may be found of $0.1\mu m^{2}$ to more than $1000 \mu m^{2}$ with non-negligible frequency. A $log\textrm{--}t$ Student' distribution $\ell t(\mu,\sigma,\nu )$ was then used because it meets both criteria and only depends on three parameters, $\mu$, $\sigma$ and $\nu$

\begin{equation}
    \ell t(A|\mu,\sigma,\nu) = \frac{\Gamma((\nu+1)/2)}{\Gamma(\nu/2)}\ \frac{1}{\sqrt{\nu \pi} \sigma\, A}\left[ 1+\frac{1}{\nu}\left(\frac{\log(A)-\mu}{\sigma}\right)^{2}\right]^{-(\nu+1)/2} \mathrm{.}
\label{eq:logt}
\end{equation}

Parameters $\mu$ and $\sigma$ describe the location and width of the distribution, respectively. Parameter
$\nu$ determines the slope of the distribution; the tails will be longer for small values of $\nu$ and shorter
for large values of $\nu$, converging on the limit of $\nu \rightarrow \infty$ to a log-normal distribution.

The complete modeling scheme consists of a $log\textrm{--}t$ distribution for areas $A$ and $B$ in the representation of circular particle measurement described in Section \ref{secCIRC}. 
Data were numerically fitted using a Markov chain Monte Carlo (MCMC) method programmed in STAN and implemented with Python's CmdStanPy library \cite{stan,cmdstanpy}. 
 

 \subsection{Probability of a false negative sample}
 

We aim to determine the conditions which may render $B = 0$ for all characteristic particles in a sample, a false negative sample (FNS).  To so so we use the full probabilistic model to infer what size values may be recorded for different pixel sizes. 

\begin{equation}
\begin{aligned}
P(\text{FNS}|\,p_{x}) &= \sum_{n=1}^{\infty} P(\{\text{all }B = 0\} |p_{x})\ P(n) \\
                                      &=\sum_{n=1}^{\infty} P(B=0|p_{x})^{n}\  P(n)
\end{aligned}
\label{eq:FN}
\end{equation}
 where:
\begin{equation}
P(B=0|\, p_{x}) = \int_{0}^{\infty} P(B=0|A,p_{x})\, P(A|D)\, dA\ .
\end{equation}.

Here, $P(n)$ is the distribution of the number of characteristic particles per sample estimated with
the frequencies observed, and $P(A|D)$ is the posterior distribution of areas $A$ integrated over the parameters fitted with the casework database $D$ (Fig.~\ref{fig:dist_muestras}).
\newpage
\section{Results}

\subsection{Circular particle acquisition model}
\begin{figure}[tbh]
\centering
\includegraphics[width=0.6\textwidth]{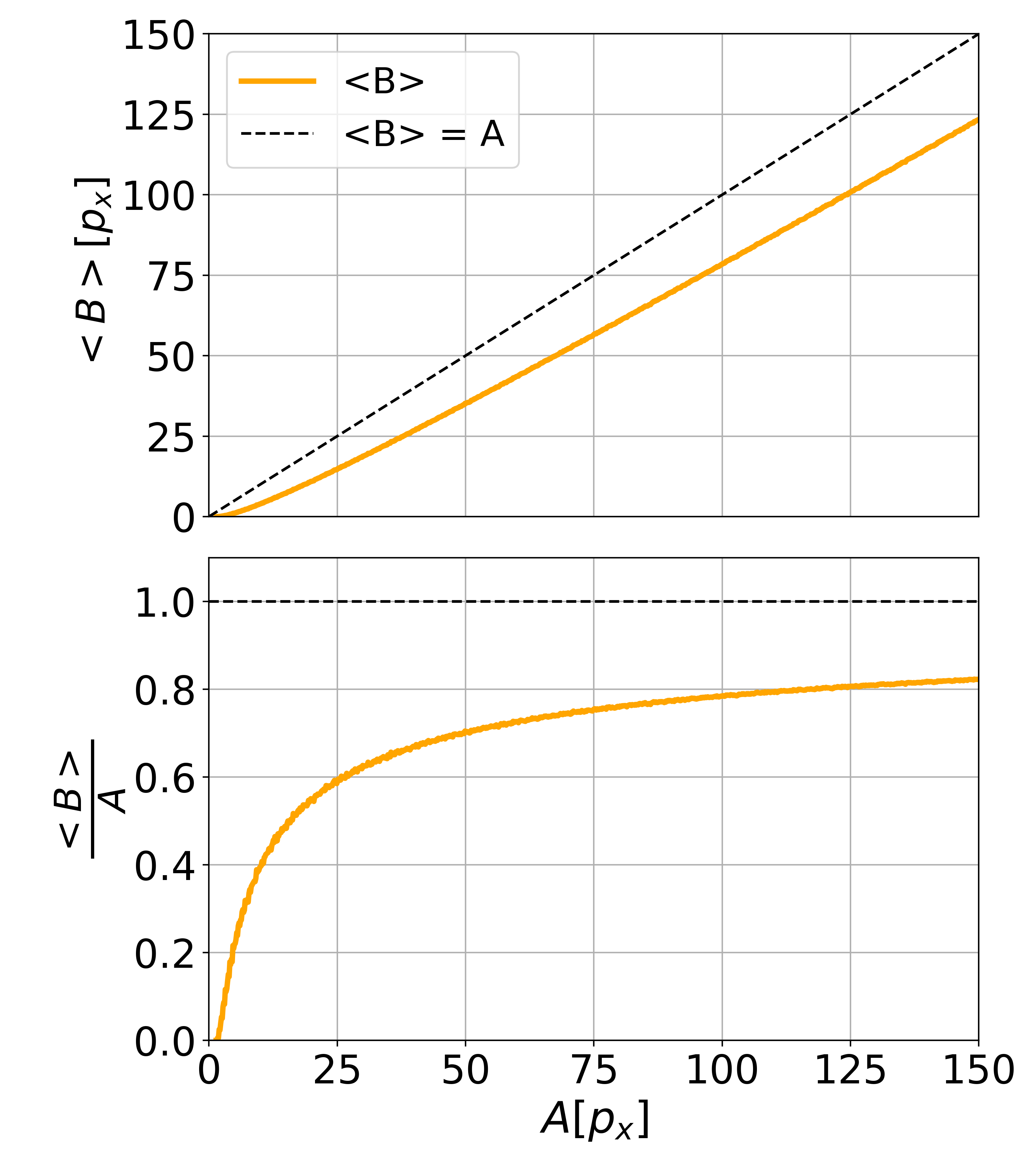}
\caption{Statistical relationship between the real area $A$ and the area $B$ observed. (Top) behavior of the
mean value of records $B$ for different values of $A$ in units of $p_{x}$.(Bottom) behavior of the quotient of the
mean value $B$ over $A$ as a function of $A$. \label{fig:B}}
\end{figure}

Under the hypothesis of circular geometry of particles we studied the statistical predictions between $B$ and $A$ which are depicted in Fig.~\ref{fig:B}. 

\begin{figure}[tbh]
\centering
\includegraphics[width=1\textwidth]{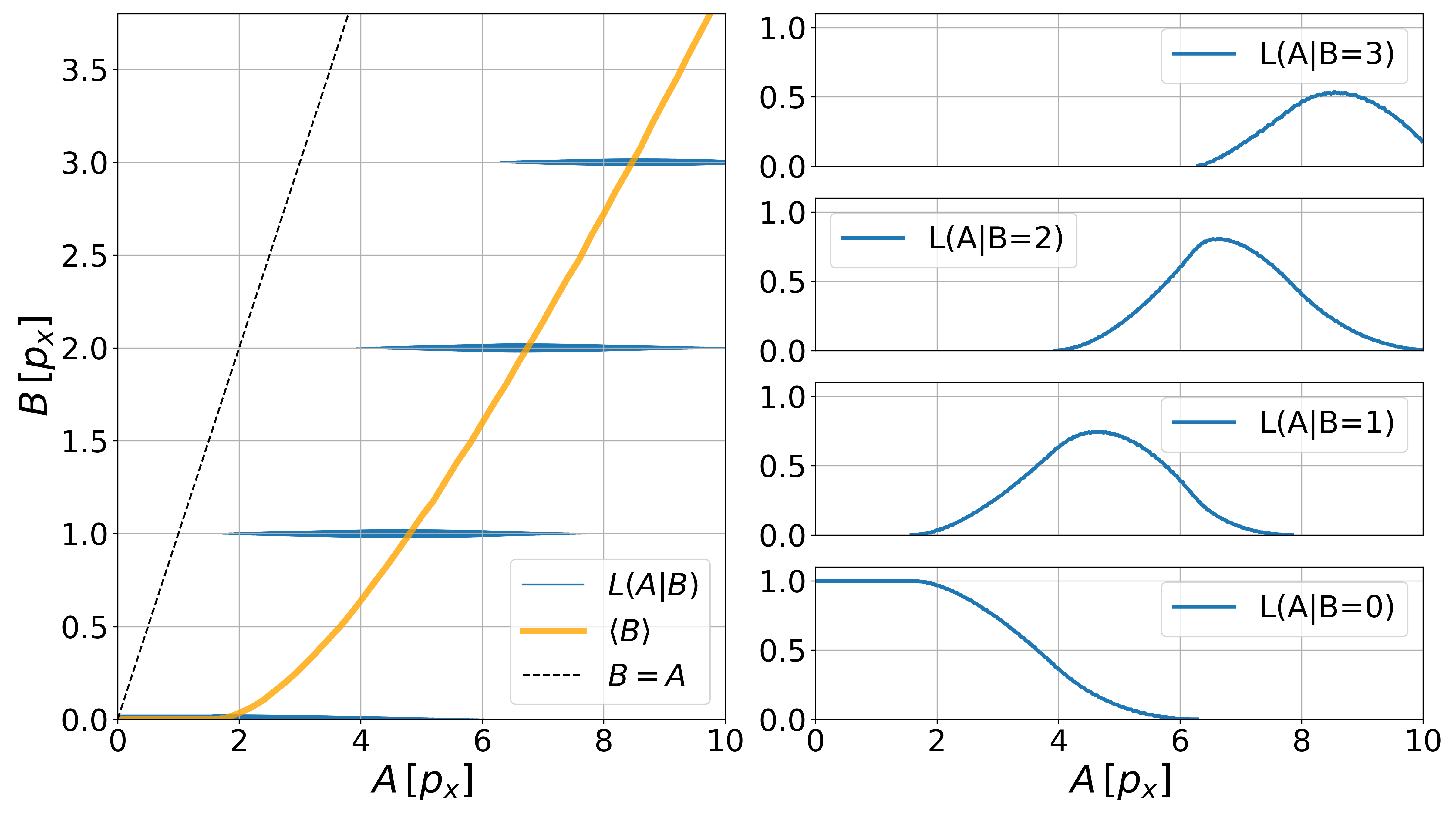}
\caption{Description of the likelihood function $L(A|B)$ obtained. a) Violin plot describing $L(A|B)$ for
different values of $B$, with wider sections for larger $L(A|B)$ values. b) Graph representing $L(A|B)$ for different values of $B$. Both graphs show a region of $A$ values which can render $B=0$, which means that a
particle may go undetected in this size range.\label{fig:LBA}}
\end{figure}

Displaying $L(A|B)$ for different values of $A$ (Fig.~\ref{fig:LBA} ) we observe that a region exists for $A\neq0$ with a probability of $B=0$, that is, some particles as large as $A=5\,p_x\,$ will not be detected. It should be noted that variable
$A$ is continuous, while $B$ only takes integer pixel values.


\subsection{Validation of the circular acquisition model}

To determine whether circular particle shape and the measurement process model are good representations of the size register, we performed successive measurements with different $p_{x}$ values on the same sample of iGSR particles. We experimentally recorded their areas and obtained a distribution of sizes $B$ which varies with $p_{x}$ but, in belonging to the same sample, is given by the same distribution of sizes $A$.

\begin{figure}[tbh]
\centering
\includegraphics[width=0.7\textwidth]{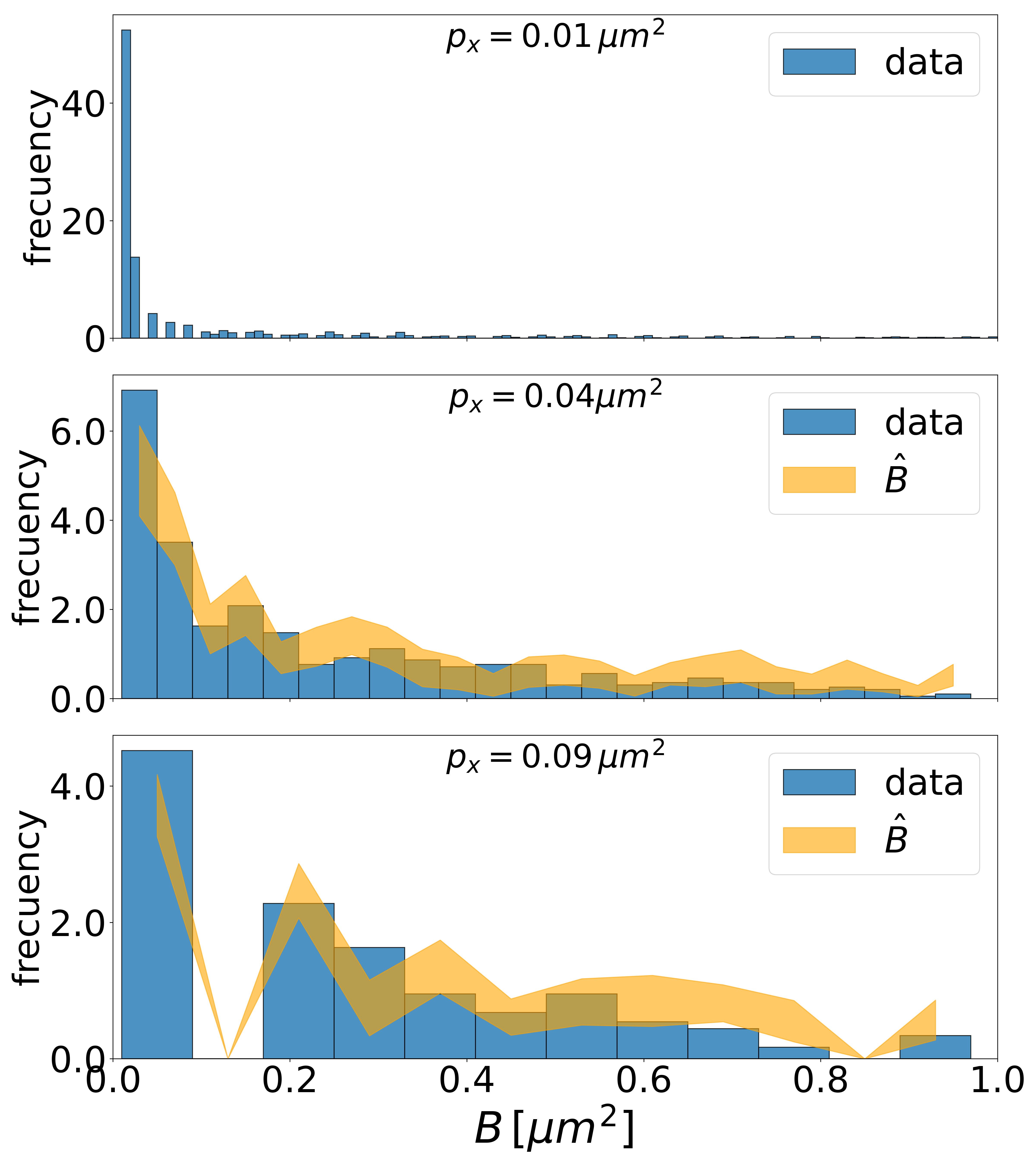}
\caption{Measurements and inferences made about the size of iGSR particles registered with different $p_{x}$ values, taking the measurements made with $p_{x}^{min} = 0.01 \mu m^2$ as the real area. For higher $p_{x}$ values, the distributions of values $B$ become wider and present gaps. Inferences in this model simultaneously reproduce both data features. The bin size of each plot is scaled to the pixel size so the relative frequencies are comparable.  \label{fig:validation}}
\end{figure}

The smallest pixel value used was close to the maximum resolution of our equipment ($p_{x}^{min} = 0.01 \mu m^{2}$). This measurement best described particle morphology in the sample and was used as an approximation of areas $A$ to infer the size distributions registered for larger pixels. To distinguish sizes registered from those inferred, $B$ will remain the notation for values registered and $\hat{B}$ will be used for values predicted by the model with the data of $p_x = 0.01\mu m^2$ as input.
\clearpage
Figure~\ref{fig:validation} presents the results of measurements for $p_{x} = (0.01; 0.04; 0.09) \mu m^2$ and inferences $\hat{B}$ for $p_{x} = (0.04; 0.09) \mu m^2$ . As $p_{x}$ grows, the distributions fall less steeply and the data show
oscillation and even gaps for some values. This may be attributed to the fact that values for particles and pixels are commensurable, as mentioned above.

The description we present here thus simultaneously reproduces these data features and validates the hypothesis of circularity and full-pixel registration as a good quantitative representation bridging real areas $A$ with measured areas $B$.

\subsection{Distribution fitting of A}

\begin{figure}[tbh]
\centering
\includegraphics[width=0.7\textwidth]{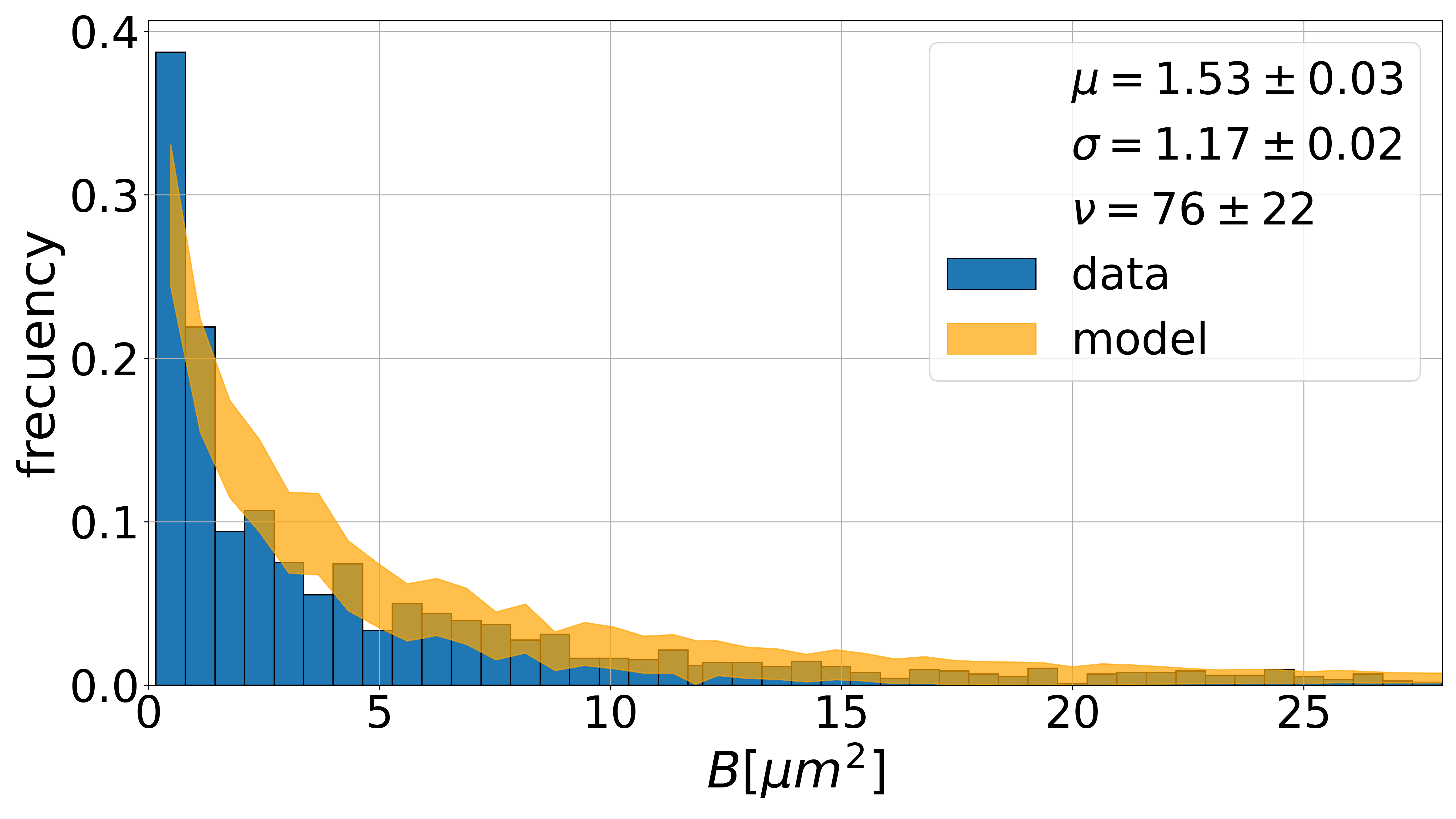}
\caption{Results of the fit of the model described in Section
\ref{sec:model} superimposed on the distribution of sizes
$B$ recorded for characteristic iGSR particles. The fitted parameters describing the log-t distribution $(\mu, \sigma, \nu)$
are indicated on the plot. Graphic evidence and parameters’ low standard deviation values strongly validate this description.
\label{fig:dist_ajuste}}
\end{figure}

In Fig.~\ref{fig:dist_ajuste} we present the fit of the full probabilistic model described in \label{sec:model} to the complete set of 2069 characteristic iGSR particle sizes recorded $B$. The parameter values obtained were:
$\mu = 1.53 \pm 0.03$, $\sigma = 1.17 \pm 0.02$ and $\nu = 76 \pm 22$. The model reproduces the data behavior of interest both for small and large particle sizes, with an explained variance $R^{2} = 0.91.$
Worth highlighting, the purpose of this modeling process is to provide a simple description of the size distribution with few parameters. Caution should be exercised in extrapolating these results to values outside the source data range $( 0.16, 1000) \mu m^2$, although this point is beyond the scope of our work.

\subsection{FNS results}

\begin{figure}[tbh]
\centering
\includegraphics[width=0.7\textwidth]{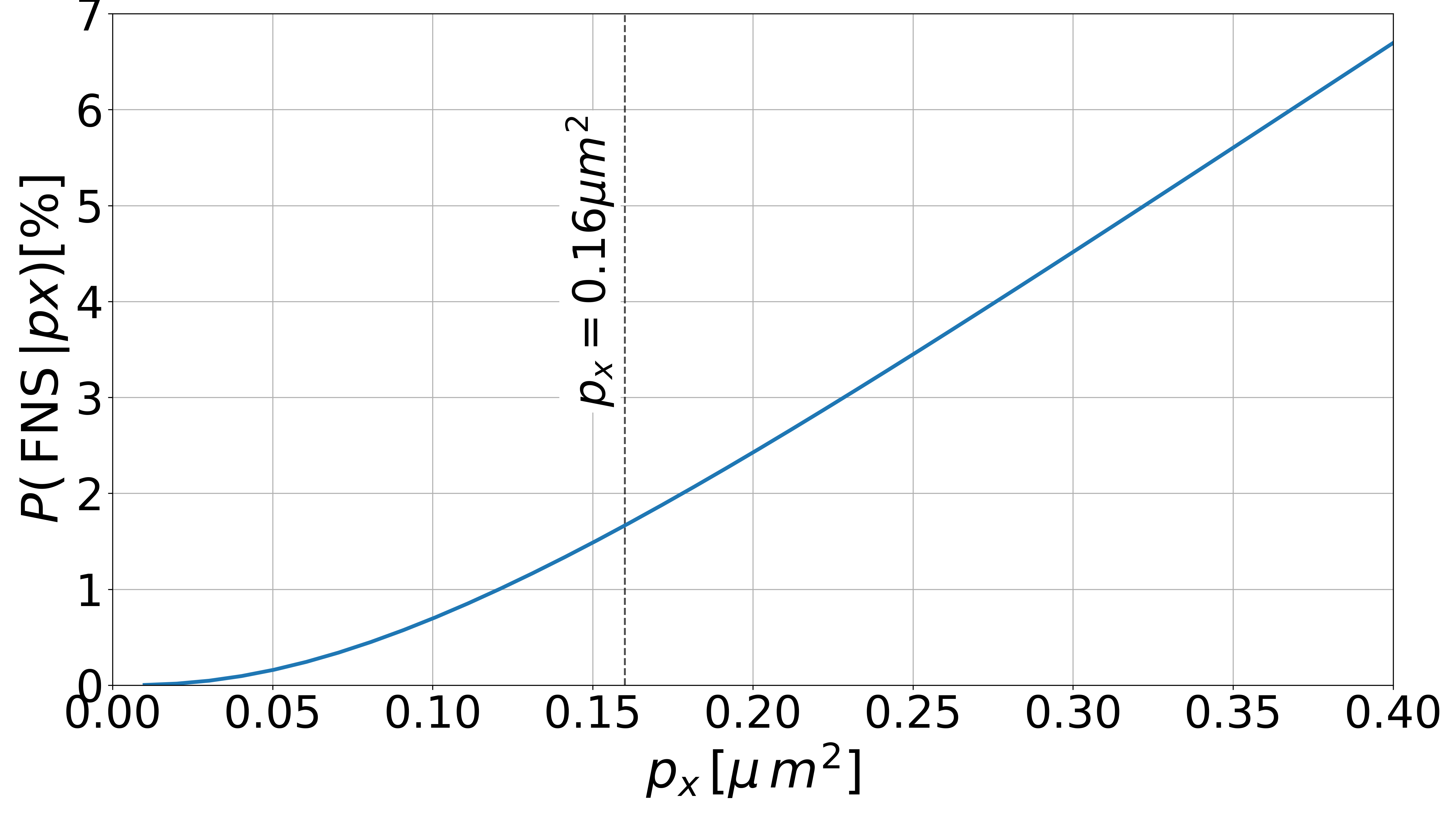}
\caption{Probability of reporting a false negative sample (FNS) results as a function of pixel size $p_{x}$. The value used for measurements $p_{x} = 0.16 \mu m^{2}$ , is indicated on the chart \label{fig:fneg}.}
\end{figure}

In order to calculate the probability of obtaining a false negative sample result we combine the parameters fitted to the distribution of $A$ with the circular particle model. In Fig.~\ref{fig:fneg} we show the probability of a FNS result in a typical iGSR sample by means of the representative particle distribution (Eq.~\ref{eq:FN}) for different values of $p_{x}$. 

As mentioned above, pixel discretization implies that some iGSR particles in the sample will inevitably fail to be detected. For this reason, any measurement of particle quantity should be expressed in probabilistic terms, and the approach presented here allows to quantify such probability, for example $P(\mathrm{X\ characteristic\ particles}\, |\,p_{x} \mathrm{\ of\ } 0.16 \mu m^{2}) = 95 \%$. 
The model predicts that only setups extremely close to $p_{x} = 0$ can guarantee that
absolutely all particles are detected. However, these are unattainable standards in everyday practice, which means that false-negative scenarios are indeed plausible and need to be described in a probabilistic manner.

In this context, the numerical results obtained here show that $P (\mathrm{FNS}| p_{x} )$ is below $7\%$ for $p_{x}$ values under $0.4 \mu m^{2}$ and $1.6\%$ for $p_{x} = 0.16 \mu m^{2}$ , the value used in data registration. These values are extremely low, as other factors (such as sample collection time) may entail particle loss in larger proportions \cite{cardinetti_proposal_2006}.

\section{Conclusions}

The sizes iGSR of particles recorded by automated systems $B$ are a statistical representation of real areas $A$. Furthermore, size measurements of the same particle may vary with its location in the sample and the pixel size $p_{x}$ used in the scan. The size measured is systematically smaller than the $A$ value for each particle, and may thus not be registered when $A$ is close to $p_{x}$. For these reasons and given that real particle sizes are not directly accessible, false negative sample results may always arise. It is in this way that the definition of $p_{x}$ regulates the probability of obtaining false negative sample results in the analysis of iGSR samples.

In this study, we used iGSR particles from 1,174 hand samples collected by the forensic science laboratory of Policía de Investigación de Rosario between 2017 and 2019. These samples realistic variability allowed us to extrapolate our results to recommendations for iGSR analysis and thus harness the laboratory’s extensive background. While we believe our results are comparable to those of other laboratories, we remain open to collaborative work to compare our findings with those of other forensic science units.

We constructed the distribution of real particle areas and calculated the probability of obtaining a false negative sample for different pixel sizes. We found that a pixel size of $0.16 \mu m^{2}$ resulted in a probability of false negative results of only $1.6\%$, which is a high effectiveness rate compared to other factors affecting the technique. We also found that pixel sizes between $0.16$ and $0.32 \mu m^{2}$ resulted in a probability of false negative results below $5\%$. These results suggest that using larger pixel sizes, which might reduce laboratory workload exponentially, could be effective in detecting characteristic particles in initial scans of samples.

Our study provides insights into the challenges of particle size measurement in iGSR analysis and highlights the importance of pixel size selection in obtaining reliable results. Our findings suggest that using larger pixel sizes can effectively reduce false negative sample results while significantly reducing laboratory workload. This preliminary work allows for the evaluation of time-saving scanning strategies and issue recommendations for measurement protocols that optimize microscope usage  and achieve more efficient and accurate iGSR analysis.

\clearpage

\printbibliography

@book{aitken,
	address = {Edinburg, UK},
	title = {Statistics and Evaluation of Evidence for Forensic Scientists},
	isbn = {0471955329},
	language = {en},
	publisher = {John Wiley \& Sons, Ltd},
	author = {Aitken, C. G. G.},
	year = {1995},
}

@article{Izraeli2014,
	title = {Optimizing {FEG}-{SEM} combined with an {SDD} {EDX} system for automated {GSR} analysis: {Optimizing} {FEG}-{SEM} for automated {GSR} analysis},
	volume = {43},
	issn = {00498246},
	shorttitle = {Optimizing {FEG}-{SEM} combined with an {SDD} {EDX} system for automated {GSR} analysis},
	url = {https://onlinelibrary.wiley.com/doi/10.1002/xrs.2495},
	doi = {10.1002/xrs.2495},
	language = {en},
	number = {1},
	urldate = {2023-02-09},
	journal = {X-Ray Spectrometry},
	author = {Izraeli, Elad S. and Tsach, Tsadok and Levin, Nadav},
	month = jan,
	year = {2014},
	pages = {29--37},
}

@article{Charles2020,
	title = {Conduction of a round-robin test on a real sample for the identification of gunshot residues by {SEM}/{EDX}},
	volume = {309},
	issn = {03790738},
	url = {https://linkinghub.elsevier.com/retrieve/pii/S0379073820300451},
	doi = {10.1016/j.forsciint.2020.110183},
	language = {en},
	urldate = {2023-02-09},
	journal = {Forensic Science International},
	author = {Charles, Sébastien and Dodier, Thierry and Kaindl, Monika and Kastéropoulos, Alain and Knijnenberg, Alwin and Larsson, Marcus and Lauper, Sandrine and Merat, Nadine and Niewoehner, Ludwig and Scholz, Thomas and Simon, Laurence},
	month = apr,
	year = {2020},
	pages = {110183},
}

@article{Ritchie2020,
	title = {Proposed practices for validating the performance of instruments used for automated inorganic gunshot residue analysis},
	volume = {20},
	issn = {24681709},
	url = {https://linkinghub.elsevier.com/retrieve/pii/S2468170920300400},
	doi = {10.1016/j.forc.2020.100252},
	language = {en},
	urldate = {2023-02-09},
	journal = {Forensic Chemistry},
	author = {Ritchie, Nicholas W.M. and DeGaetano, Doug and Edwards, Dave and Niewoehner, Ludwig and Platek, Frank and Wyatt, J. Matney},
	month = aug,
	year = {2020},
	pages = {100252},
}

@inproceedings{White2014,
	address = {Monterey, California, United States},
	title = {Developing a quality assurance program for gunshot primer residue analysis},
	url = {http://proceedings.spiedigitallibrary.org/proceeding.aspx?doi=10.1117/12.2073770},
	doi = {10.1117/12.2073770},
	language = {en},
	urldate = {2023-02-09},
	author = {White, Thomas R.},
	editor = {Postek, Michael T. and Newbury, Dale E. and Platek, S. Frank and Maugel, Tim K.},
	month = sep,
	year = {2014},
	pages = {92360K},
}

@inproceedings{Charles2010,
	address = {Monterey, California},
	title = {The {Bayesian} approach to reporting {GSR} analysis results: some first-hand experiences},
	shorttitle = {The {Bayesian} approach to reporting {GSR} analysis results},
	url = {http://proceedings.spiedigitallibrary.org/proceeding.aspx?doi=10.1117/12.853446},
	doi = {10.1117/12.853446},
	language = {en},
	urldate = {2023-02-17},
	author = {Charles, Sebastien and Nys, Bart},
	editor = {Postek, Michael T. and Newbury, Dale E. and Platek, S. Frank and Joy, David C.},
	month = jun,
	year = {2010},
	pages = {77291B},
}

@book{ASTM2020,
	title = {Standard Practice for Gunshot Residue Analysis by Scanning Electron Microscopy/Energy Dispersive X-Ray Spectrometry},
	language = {en},
	institution = {ASTM International},
	author = {{ASTM-International}},
	publisher = {Designation: E1588 20 (2020)},
	doi = {10.1520/E1588-20},
}

@article{WoltenI,
	title = {Particle analysis for the detection of gunshot residue. I: Scanning electron microscopy/energy dispersive X-ray characterisation of hand deposits from firing},
	volume = {24},
	language = {en},
	number = {2},
	journal = {Journal of Forensic Sciences},
	author = {Wolten, G.M. and Nesbitt,  R.S. and Calloway, A.R. and  Loper, G.L. and Jones, P.F.},
	year = {1979},
}

@book{Gelman,
  author = {Gelman, Andrew and Carlin, John B. and Stern, Hal S. and Rubin, Donald B.},
  edition = {3rd ed.},
  publisher = {Chapman and Hall/CRC},
  title = {Bayesian Data Analysis},
  year = 2004
}

@book{taroni,
	address = {Chichester, UK},
	title = {Data {Analysis} in {Forensic} {Science}: {A} {Bayesian} {Decision} {Perspective}},
	isbn = {978-0-470-66508-4 978-0-470-99835-9},
	shorttitle = {Data {Analysis} in {Forensic} {Science}},
	language = {en},
	urldate = {2020-11-04},
	publisher = {John Wiley \& Sons, Ltd},
	author = {Taroni, Franco and Bozza, Silvia and Biedermann, Alex and Garbolino, Paolo and Aitken, Colin},
	month = apr,
	year = {2010},
	doi = {10.1002/9780470665084}
}

@article{zeichner_more_1997,
	title = {More on the {Uniqueness} of {Gunshot} {Residue} ({GSR}) {Particles}},
	volume = {42},
	issn = {00221198},
	url = {http://www.astm.org/doiLink.cgi?JFS14255J},
	doi = {10.1520/JFS14255J},
	language = {en},
	number = {6},
	urldate = {2021-08-10},
	journal = {Journal of Forensic Sciences},
	author = {Zeichner, Arie and Levin, Nadav},
	month = nov,
	year = {1997},
	pages = {14255J},
}

@manual{stan,
author = {Stan Development Team},
title = {Stan Modeling Language Users Guide and Reference Manual},
language = {English},
version = {Version 2.29},
url = {https://mc-stan.org},
}

@manual{cmdstanpy,
author = {Stan Development Team},
title = {CmdStanPy},
language = {English},
version = {Version 1.01},
url = {https://mc-stan.org/cmdstanpy/},
}

@article{cardinetti_proposal_2006,
	title = {A proposal for statistical evaluation of the detection of gunshot residues on a suspect},
	volume = {28},
	issn = {01610457, 19328745},
	url = {https://onlinelibrary.wiley.com/doi/10.1002/sca.4950280302},
	doi = {10.1002/sca.4950280302},
	language = {en},
	number = {3},
	urldate = {2021-08-10},
	journal = {Scanning},
	author = {Cardinetti, Bruno and Ciampini, Claudio and Abate, Sergio and Marchetti, Christian and Ferrari, Francesco and Di Tullio, Donatello and D'onofrio, Carlo and Orlando, Giovanni and Gravina, Luciano and Torresi, Luca and Saporita, Giuseppe},
	month = dec,
	year = {2006},
	pages = {142--147},
}

@article{Timan1987,
	title = {Automated {Gunshot} {Residue} {Particle} {Search} and {Characterization}},
	volume = {32},
	issn = {00221198},
	url = {http://www.astm.org/doiLink.cgi?JFS12327J},
	doi = {10.1520/JFS12327J},
	language = {en},
	number = {1},
	urldate = {2019-02-11},
	journal = {Journal of Forensic Sciences},
	author = {Tillman, Warren L.},
	month = jan,
	year = {1987},
	pages = {12327J},
}

@article{Kee1987,
	title = {Casework assessment of an automated scanning electron microscope/microanalysis system for the detection of firearms discharge particles},
	volume = {27},
	issn = {00157368},
	url = {https://linkinghub.elsevier.com/retrieve/pii/S0015736887727716},
	doi = {10.1016/S0015-7368(87)72771-6},
	language = {en},
	number = {5},
	urldate = {2022-06-07},
	journal = {Journal of the Forensic Science Society},
	author = {Kee, T.G. and Beck, C.},
	month = sep,
	year = {1987},
	pages = {321--330},
}

\end{document}